\documentclass{article}

\usepackage{graphicx}%
\usepackage{multirow}%
\usepackage{amsmath,amssymb,amsfonts}%
\usepackage{amsthm}%
\usepackage{mathrsfs}%
\usepackage[title]{appendix}%
\usepackage{xcolor}%
\usepackage{textcomp}%
\usepackage{manyfoot}%
\usepackage{booktabs}%
\usepackage{algorithm}%
\usepackage{algorithmicx}%
\usepackage{algpseudocode}%
\usepackage{listings}%
\usepackage{array}
\usepackage{subcaption}
\usepackage{url}
\usepackage{eurosym}
\usepackage{authblk}

\usepackage{natbib}
\algnewcommand\algorithmicforeach{\textbf{for each}}
\algdef{S}[FOR]{ForEach}[1]{\algorithmicforeach\ #1\ \algorithmicdo}

\theoremstyle{thmstyleone}%
\theoremstyle{thmstyletwo}%

\theoremstyle{thmstylethree}%

\usepackage{todonotes}

\usepackage{comment}

\title{Mobility to Campus - a Framework to Evaluate and Compare Different Mobility Modes}

\author{Helena Fehler}
\author{Marco Pruckner}
\author{Marie Schmidt}

\affil{Department for Computer Science, University of Würzburg, Am Hubland, 97074 Würzburg, Germany, marie.schmidt@uni-wuerzburg.de  marco.pruckner@uni-wuerzburg.de}

\date{}

\begin{document}
\maketitle

\abstract{
The transport sector accounts for about 20\% of German CO$_2$ emissions, with commuter traffic contributing a significant part.
Particularly in rural areas, where public transport is inconvenient to use, private cars are a common choice for commuting and most commuters travel alone in their cars.
Consolidation of some of these trips has the potential to decrease CO$_2$ emissions and could be achieved, e.g., by offering \emph{ridesharing} (commuters with similar origin-destination pairs share a car) or \emph{ridepooling} (commuters are picked up by shuttle services).
In this study, we present a framework to assess the potential of introducing new mobility modes like \emph{ridesharing} and \emph{ridepooling} for commuting towards several locations in close vicinity to each other.

We test our framework on the case of student mobility to the University of Würzburg, a university with several campus locations and a big and rather rural catchment area, where existing public transport options are inconvenient and many students commute by car. 
We combine data on student home addresses and campus visitation times to create demand scenarios.
In our case study, we compare the mobility modes of \emph{ridesharing} and \emph{ridepooling} to the base case, where students travel by car on their own.
We find that \emph{ridesharing} has the potential to greatly reduce emissions, depending on the percentage of students willing to use the service and their willingness to walk to the departure location. The benefit of \emph{ridepooling} is less clear, materializing only if the shuttle vehicles are more energy efficient than the student cars.
}
\bigskip


\maketitle

\section{Introduction}\label{sec1}
In 2021, $15\%$ of greenhouse gas emissions worldwide were caused by transport and mobility \cite{owid-emissions-by-sector}.
According to research institute \emph{Agora Verkehrswende} nearly a quarter of all mobility emissions in Germany are caused by commuting~\cite{pendeln}, which includes both employee and (university) student commuting.

Although German university students typically receive annual tickets for public transport,
according to a study from 2018~\cite{hochschule} about $25\%$ of them commute in their private cars. 
The study also reveals that this percentage varies widely between urban and rural regions.
While only $18\%$ of students from urban areas travel by car, the percentage rises to $41\%$ for rural areas~\cite{mid}, which can be explained by bad public transport service there \cite{sorensen2021much}.

Organizing the commuting to campus comes with some additional difficulties compared to standard 'commuting to work'-scenarios.
Unlike employees in some industries, for whom a fixed start time is still the standard, university students have varying class start times that often change daily throughout the week.
Moreover, their preferred commuting times do not have to correlate with their classes since students oftentimes stay at their campus for other activities such as studying, socializing and eating at the cafeteria. 
However, also for employees, non-standard schedules are becoming more common \cite{bolino2021working}.
Shared mobility is often introduced in hope of pooling demand and thereby reducing emissions, while still providing a sufficient level of service to be an attractive alternative to travel by own car \cite{sorensen2021much,santos2018sustainability,wang2022assessing}.

In this paper we propose a framework that is open-source\footnote{https://github.com/H-Fehler/CaMoF} and allows to compare different mobility modes
for scenarios that model commuting to a central campus with one or several 
campus locations in close vicinity to each other.
We describe framework and findings in terms of student commuting, but remark that it is also applicable to other mobility scenarios, in which commuters have a central campus, possibly with multiple on-campus locations, as destination.

We have implemented and tested two mobility modes in the framework: \emph{ridesharing} and \emph{ridepooling}.
In \emph{ridesharing}, private car owners, when planning a ride, offer to take passengers along for the entire ride or part of it \cite{lukesch2019sharing}.
In \emph{ridepooling}, vehicles, in most cases owned by a \emph{ridepooling} company, pick up passengers at their origins and bring them to their destinations \cite{vansteenwegen2022survey}. 
The main difference between the two modes thus lies in who owns and operates the vehicles. This leads to differences in vehicle characteristics, most notably number of seats and energy efficiency, between the modes. It also has a direct impact on operational characteristics, in particular flexibility of routes, pickup locations and the willingness to accept detours from the direct route. There may also be an impact on trust, and willingness to use the mobility mode at all, compare \cite{werth2021examining,bachmann2018drives,olsson2019they}, which we consider out of scope for our study.
We further disregard road traffic in the routing functionalities offered by the framework.

The remainder of the paper is organized as follows: In Chapter~\ref{sec:relatedwork} we present an overview on works on \emph{ridesharing} and \emph{ridepooling} as well as simulation studies on mobility.
In Chapter~\ref{sec:methodology} we describe our framework architecture, method of demand generation and implemented mobility modes.
Chapter~\ref{sec:case_study} contains a description of our case study and its results and we conclude this work in Chapter~\ref{sec:conclusion} with a discussion of the results and opportunities for future research.

\section{Related work}\label{sec:relatedwork}

The pressure to reduce mobility-related CO$_2$ emissions, as well as the rise of mobile technology that allows fast communication and flexible interaction between vehicles and users, has led to a surge in interest in innovative mobility modes, such as \emph{ridepooling} and \emph{ridesharing}, which could replace or supplement more traditional modes like driving in (self-owned) cars, public transport, walking and cycling.

One approach to gain further knowledge on this topic is by conducting surveys and interviews with prospective users in order to gain insights about potential usage and sensitivity to key service parameters.
For instance, Petrik et al.~\cite{petrik2018transport} conduct a survey among focus groups in  Auckland~(New Zealand), Dublin~(Ireland) and Helsinki~(Finland) regarding influence factors on acceptance of mobility modes. They find that cost, travel time and accessibility, i.e.~waiting time and distance to boarding location, are the most important factors.

A second body of literature related to this topic addresses the algorithmic challenges of matching passengers for \emph{ridesharing} trips and of routing and scheduling \emph{ridepooling} vehicles.
Vansteenwegen et al.~\cite{vansteenwegen2022survey} provide a comprehensive literature overview and classification of \emph{ridepooling} systems and algorithmic approaches for routing and scheduling the respective vehicles. 
A survey on ride-matching algorithms is given in Tafreshian et al.\cite{tafreshian2020frontiers}.
According to \cite{vansteenwegen2022survey,tafreshian2020frontiers} typical goal functions consider passenger-centric \emph{quality of service} metrics such as waiting times, travel times, and passenger costs, and operator-related objectives such as vehicle travel time and distance, as well as operational costs. 
Another common evaluation function is to measure environmental impacts such as CO$_2$ emissions.
For example, Liyanage and Dia~\cite{liyanage2020agent} develop an agent-based simulation in which they compare on-demand public transport to scheduled bus services in Melbourne, Australia, regarding emissions among other metrics.

Simulation is a common approach to consider and evaluate different mobility modes in comparison to each other.
An overview of commercial and open-source traffic simulation tools with a focus on agent-based tools is given by Nguyen et al.~\cite{nguyen2021overview}.
Huan et al.~\cite{huang2022overview} give an overview on agent-based models which they categorize based on temporal scope, study objective, and types of agents included, also discussing the strengths and weaknesses of the respective models.
Another common use case for the application of simulations is comparing the effects of different adoption rates.
For example, Bistaffa et al.~\cite{bistaffa2019computational} assess multiple adoption rates for \emph{ridesharing} in relation to quality of service and sustainability metrics. They find that a 20\% \emph{ridesharing} adoption rate is required to achieve a 10\% reduction in CO$_2$ emissions, while an 80\% adoption rate is necessary for a 50\% reduction in emissions.

While several studies examine mobility opportunities within cities as alternatives to services like taxis, our focus is on students who reside outside their university city and undertake long-distance commutes, as opposed to short-distance urban trips. Liu et al.~\cite{liu2019bus} present a study that, similar to our work, concentrates on individuals commuting from outside a city. They investigate a \emph{ridepooling} service (which they name bus \emph{ridesharing}) where passengers request long-distance trips and wait until a sufficient amount of people gather for the ride.
The authors focus on optimizing ride-matching by addressing the capacitated clustering problem of travel demand and the location-allocation problem for pick-ups and drop-offs. Afterward, they execute constraint based pruning.
They develop both exact and approximate algorithms for this purpose.
To demonstrate the efficiency of the proposed service, they utilize a real-life dataset from Shanghai taxis.
Their results indicate that the service is both cost and energy efficient.

The majority of \emph{ridepooling} studies utilize autonomous vehicles and avoid operational constraints in that way. However, since autonomous vehicles are not yet widely used, examining \emph{ridepooling} with professional drivers would result in more realistic evaluations. 
Zwick et al.~\cite{zwick2022shifts} executed a simulation and evaluation of a non-autonomous \emph{ridepooling} approach, comparing it to autonomous \emph{ridepooling}. Their results show that non-autonomous \emph{ridepooling} is significantly constrained  by operational factors which limit the number of served rides. Moreover, they demonstrate that shift plans substantially influence these outcomes which stresses the importance of considering human factors in \emph{ridepooling} systems.

To the best of our abilities, we could not find any works in the literature that compare \emph{ridesharing} and \emph{ridepooling} for campus mobility.
Our work further contributes a framework for modeling and evaluating multiple mobility modes for campus mobility under identical circumstances and hard constraints. This framework allows for easy and dynamic adjustment of input parameters and the addition of new mobility modes. We also develop and evaluate our own heuristic-based approaches for \emph{ridesharing} and \emph{ridepooling} and assess which input parameters are critical for the results of each mobility mode.

\section{Methodology}\label{sec:methodology}

In the following we describe our decision-support framework, which is open-source and allows to evaluate and compare various mobility modes on different demand scenarios and is described in more detail in Section~\ref{sec-framework}. 
Furthermore, we have implemented three mobility modes, \emph{Ridepooling}, \emph{Ridesharing}, and the baseline mode \emph{EverybodyDrives} which are described in Section~\ref{sec-mobilitymodes}. Other mobility modes can be added by implementing the necessary interfaces.

Our framework is applicable to scenarios that model mobility towards a central region, which may consist of one location or several locations in close proximity.
In Section~\ref{sec-demand} we detail our approach to generate such a scenario from student home location and travel data.
However, our framework is not restricted to demand scenarios generated as described in Section~\ref{sec-demand}, but can work on any demand scenario that specifies agent home locations and travel-time requests.

\subsection{Framework}\label{sec-framework}
Our framework is implemented in Java, an object-oriented programming language, and provides objects representing components such as requests, agents, rides, vehicles, and events.

\texttt{Agent} objects represent people traveling to and from campus, containing information about their home location, owned vehicle, and travel requests. \texttt{Request} objects detail daily travel demands, including preferred arrival and departure times, submission time, and parking coordinates near campus. \texttt{Vehicle} objects store information about available seats, driver, fuel consumption, and associated costs and emissions, enabling ride-specific calculations.
\texttt{Match} objects compile information about potential groupings of agents with feasible requests.
\texttt{Ride} objects represent completed journeys with zero, one, or multiple agents, which are either headed to campus or home, or can cover both directions (simultaneously transporting an agent for a home drop-off and an agent for a campus drop-off).
These objects include information about the driver, participating agents, ride start and end times and coordinates, as well as information about stops made (such as location and reason for stopping).
Additionally, \texttt{Event} objects represent moments in time when incidents such as a ride start occur which is useful for modes implemented as simulations.

The framework is configured by a number of configuration parameters, including global parameters~(e.g.~maximum distance to campus) and mode-specific parameters~(e.g.~number of available \emph{ridepooling} shuttle buses).
A key input for the framework configuration is the file containing a demand scenario which is used to build \texttt{Agent} objects. The framework then invokes the specified mobility modes to generate rides for the agents and compute performance metrics.

\begin{figure}
    \includegraphics[page=1,width=\textwidth]{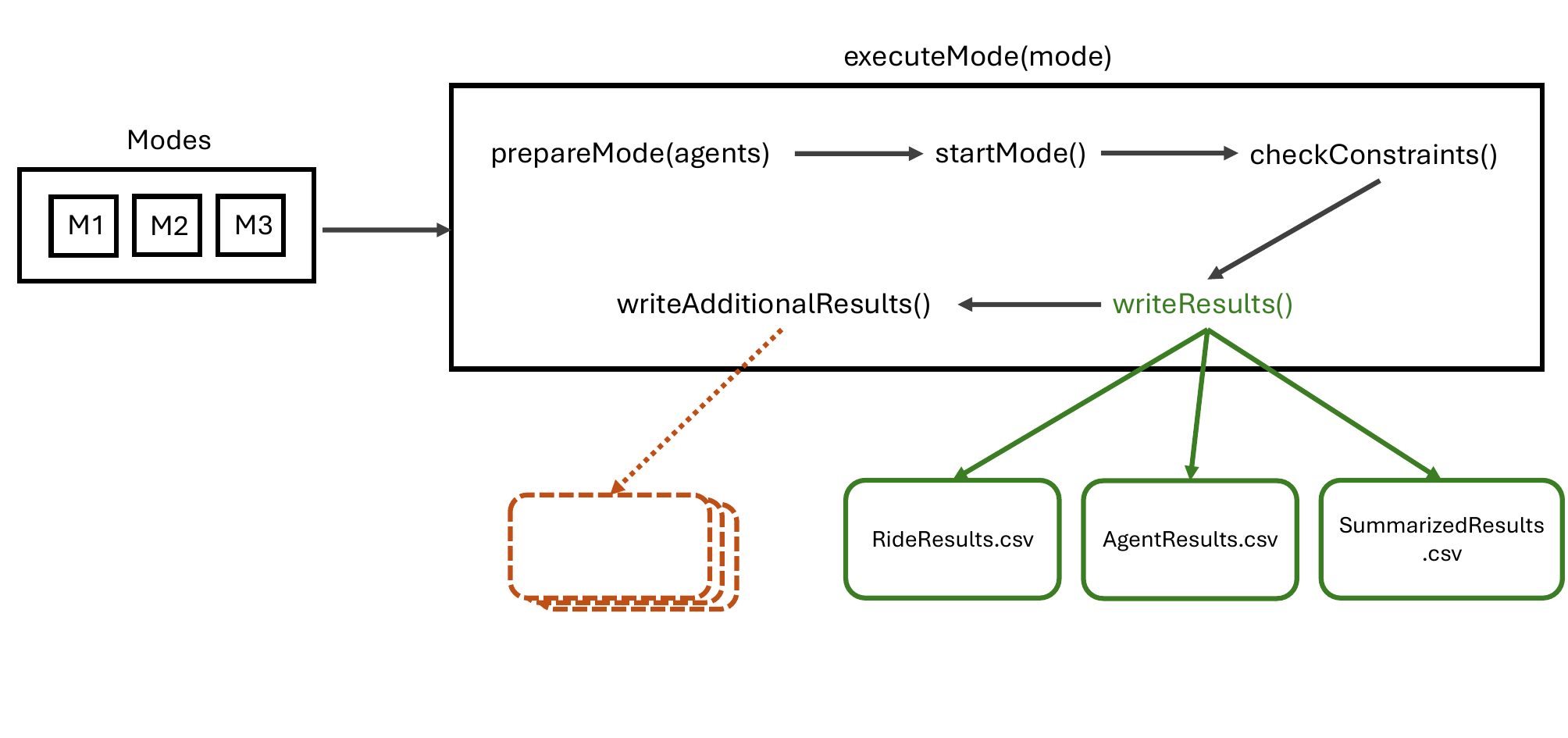}
        \caption{The framework's mode execution process in which the method $executeMode$ is started for each mode.}
        \label{fig:mode_execution}
\end{figure}

Added modes should follow the framework's template for mobility modes which prescribes interfaces for inputting agents, executing the mode and returning output metrics.
Figure~\ref{fig:mode_execution} depicts our framework's process of mode execution. Each chosen mode is input into the framework’s \texttt{executeMode} method which utilizes the mode’s interfaces. First, it calls the \texttt{prepareMode} method, using agent objects as input. This method can be implemented to, for example, execute calculations which are necessary for the mode’s execution which is then started with the interface method \texttt{startMode}. Following this, the framework calls upon the interface method \texttt{checkConstraints} which can be implemented to verify adherence to all agents' constraints or detect erroneous mode execution. Lastly, the framework executes the mode’s methods \texttt{writeResults} and \texttt{writeAdditionalResults}.

The general output metrics are CO$_2$ emissions~(g), cost~(\euro), traveled distance, and time.
These metrics are given aggregated over the day, but also disaggregated per agent and per ride. 
The method \texttt{writeResults} is already implemented by the framework itself and produces the three csv files \texttt{rideResults.csv}~(containing each ride with its corresponding metrics), \texttt{agentResults.csv}~(containing each agent's metrics) and \texttt{summarizedResults.csv}~(containing total metric values for the executed day).
Additionally, mode-specific metrics can be output by a mode if needed.
This outputting should be implemented in the method \texttt{writeAdditionalResults}.
Additionally, when a new mode is added, its corresponding mode specific parameters have to be added to the configuration.
By including such parameters, we are able to evaluate different scenarios for each mobility mode dynamically and fairly easily.

The framework provides several further functionalities (in the following called \emph{helper functions}) which can be called upon to support in the ride generation, such as
routing using the tool \emph{Graphhopper}~\cite{graphhopper} and traveling salesperson problem (TSP) optimization with the help of the tool \emph{jsprit}~\cite{jsprit} as well as the evaluation of rides with respect to the general output metrics which can be used in the modes.
The open-source routing tool \emph{Graphhopper} takes the attributes of road stretches into account which influence the travel time, such as road type or elevation changes. However, \emph{Graphhopper} alone does not factor in road traffic.

By default our framework contains one demand scenario for the University of Würzburg and three mobility modes along with their corresponding output metrics. These are described in Sections~\ref{sec-demand} and~\ref{sec-mobilitymodes}, respectively.

\subsection{Implemented Mobility Modes}\label{sec-mobilitymodes}
We supply three default mobility modes with our framework.
In the first mode, \emph{EverybodyDrives}, all agents travel by privately owned car.
This mode can be used as a baseline to compare other modes to.
The two other modes, which aim to serve as possibly more sustainable campus mobility alternatives, 
are \emph{Ridesharing} and \emph{Ridepooling}.

The three implemented mobility modes have a few properties in common which we list in the following. Although used in the three default mobility modes, these properties are not crucial for using a mode in the framework and do thus not impose restrictions on newly added modes:

Firstly, all agents prefer to use the mode being considered and do so if possible.
We make this assumption in order to evaluate a best hypothetical scenario.
For the same reason, we assume that all agents own a private car, so that they have a fallback option if they cannot be transported with \emph{Ridepooling} or \emph{Ridesharing}.
Thirdly, for both \emph{Ridesharing} and \emph{Ridepooling} we assume that requests arrive one after the other over the day in an online fashion. That is, our modes do not compute a globally optimal solution based on the full information available in the demand scenario, but model an online approach based on sequential revelation of requests.
Lastly, since our focus is not on the influence of traffic on \emph{Ridepooling} and \emph{Ridesharing}, we do not use a microscopic traffic model in this work. Instead, we focus on the feasibility and the effects of different mobility
modes on metrics like emissions and costs under the same circumstances.
Therefore, we also do not consider agents’ choice behavior since the different scenarios are evaluated
separately.

While our three included modes are implemented as dynamic, discrete-event agent-based simulations (in which we process online incoming requests sequentially for \emph{Ridesharing} and \emph{Ridepooling}), newly added modes can compute rides and general output metrics in different ways.

\subsubsection{EverybodyDrives}
The \emph{EverybodyDrives} mode simulates a scenario in which each agent uses his or her personal vehicle for campus commutes. After receiving the input travel requests, the mode extracts each agent's preferred arrival and departure times and locations. It then calculates the required departure time for each agent to arrive punctually at campus. The route is determined by finding the shortest path based on deterministic, i.e. non-traffic dependent, travel time for which we use the tool Graphhopper, which is provided by the framework's helper functions.

\subsubsection{Ridesharing}
For our \emph{Ridesharing} mode we assume that all agents are willing to serve as drivers using their personal vehicles.
Drivers depart from their homes and drive directly to the university campus. However,
they are willing to drop off passengers at different campus locations as long as this
does not substantially increase their travel time.
Passengers, i.e.~agents that are not designated as drivers, are willing to walk up to a certain distance to and from the driver's home location before and after the trip.
We further assume that all agents are
willing to accept some flexibility with their preferred arrival and departure times.
We define arrival and departure time windows $W_A=[T_A-FM;T_A+FM]$ and $W_D=[T_D;T_D+2FM]$) using a \texttt{flexible time} (FM) input parameter which is measured in minutes and applied to the preferred arrival time ($T_A$) or departure time ($T_D$), respectively.
This allows accommodating mobility modes that lead to (small) detours with respect to the passengers' direct routes.

Our \emph{Ridesharing} approach models independent arrival and departure requests, which are then sequentially matched throughout the day. We choose separating the requests since more home-bound ride opportunities emerge as the day progresses which agents do not know about at the start of their day. 
As arrival and departure requests are handled separately, a passenger may not be able to find a return trip. We refer to these passengers as lost passengers.
A lost passenger could utilize an alternative way of coming home, either by using public transportation or multi-station carsharing. We consider both options in our evaluation.

Our \emph{Ridesharing} mode is implemented as a discrete-event simulation. Initially, a list of events is created based on input requests, sorted by request submission time. For this, the framework separates combined departure and arrival requests, as provided
by the framework, and processes them sequentially.
Each resulting departure request is assigned a new submission time, calculated by subtracting a randomly generated number, ranging from 30 to 120 minutes, from the agent's preferred departure time from campus.

For each request, we assess its potential assignment to a match of the same type (arrival or departure).
For this we differentiate between arrival and departure requests. For an arrival request we begin by sorting all existing matches of the corresponding type based on their similarity to the request. This similarity is determined by the proximity of time windows and home locations. We then evaluate the feasibility of assigning the request to each of the 20 (or fewer) most similar matches.

We verify the feasibility of an assignment by checking if
\begin{itemize}
    \item the driver’s home position is inside the specified accepted walking distance to the request's corresponding agent,
    \item no waiting times occur during the resulting ride,
    \item there exists a route in which said ride's agents arrive/depart during their time window, and
    \item all of said ride's agents spent no more minutes traveling in a vehicle than their accepted \emph{Ridesharing} travel time.
\end{itemize}

These feasibility calculations are performed using our helper functions for routing and TSP calculations.
With this the \emph{Ridesharing} mode checks if it is possible to serve all needed stops in their respective time windows and calculates the optimal path.
For the sake of efficiency we first verify whether an overlap between the request time window and an already existing time window of a stop of the match exists before applying the TSP helper.
Afterwards, out of all feasible matches we assign the request to the match in which the agent has to walk the smallest distance to or from the driver.
This match is then updated by adding the new agent to the passenger list and recalculating the necessary ride start time of this match.
Until this start time is exceeded further agents can join the match as long as the car is not fully occupied.

In the absence of a fitting match for the current campus-bound request, the request agent takes on the role of driver, offering to transport other passengers. 
This action creates a new \emph{Ridesharing} match, increasing the number of available \emph{Ridesharing} options.

In the case of a departure request we differentiate between agents who were the driver in their match to campus and agents who are dependent on finding a match home.
Agents who were the driver before are also assigned as drivers for the return trip whereas the latter undergo the aforementioned match assignment process. However, if no fitting match is found, these agents are added to a waiting list that is also considered for subsequent matching.
However, as soon as the simulation's timestamp, i.e.~the timestamp of the event being processed currently, exceeds their departure time window, these agents are denoted as lost.

For the lost passengers, we add a penalty to the performance indicator \emph{costs} and in addition, report their number as an additional output metric.
In our experiments we have chosen not to assign penalties to the  \emph{traveled distance}, \emph{CO$_2$ emissions} and \emph{traveled time}, instead using the lost student's usual direct travel information from campus to home.
This would be in line with the assumption that these passengers use multi-station carsharing which is convenient but more expensive.

The mode then uses the framework's helper functions for output evaluations to compute the general output metrics \emph{traveled distance}, \emph{CO$_2$ emissions}, \emph{costs} and \emph{traveled time}.
Additionally, the mode-specific output metrics \emph{walking time} and \emph{walking distance} as well as \emph{lost passengers} are output by the \emph{Ridesharing} mode.

\subsubsection{Ridepooling}

In our \emph{Ridepooling} mode, passengers are picked up from their homes and transported to various campus locations by an on-demand shuttle service. This service uses a fleet of autonomous minibuses.
We use the same time windows for departure and arrival times as those utilized in our \emph{Ridesharing} mode. Unlike for \emph{Ridesharing}, where passengers are grouped at one stop, this shuttle service picks up and drops off each passenger directly at their home.
Furthermore, a passenger only uses \emph{Ridepooling} if both the trip to campus and the return trip can be guaranteed through this service. If either trip cannot be assured, the passenger opts to use their own car for transportation.
Thus, our mode models a situation in which travel requests must be submitted simultaneously for both directions, unlike in our \emph{Ridesharing} mode.
These requests are then handled as a whole, rather than as separate departure and arrival requests.
The mode depends on a vehicle fleet whose exact number and vehicle specific attributes can be set with mode-specific configuration parameters.
Vehicles are stored in depots, which have configurable locations and capacities. Each vehicle starts its first service from its depot and returns to the nearest available depot after its last serviced ride. However, the vehicles remain at the last drop-off location between serviced rides instead of returning to depots.

Our \emph{Ridepooling} mode is implemented as a discrete-event simulation. Initially, a list of events is created  from the requests, sorted by submission times. Unlike \emph{Ridesharing}, these requests are processed as a whole, taking both trip directions into account.
The requests are then matched sequentially throughout the day, as detailed below.
In order to limit the size of the list of matches that are to be evaluated, we partition the serviced region into a configurable number of circle segments of same angle with the centroid of the different campus locations at the center. 
Requests are assigned to segments based on the agent's home position, with an exception for agents residing in close proximity to their campus destination (defined by a configurable distance parameter).
This allows us to reduce the amount of examined matches by only considering matching requests of the same or bordering segments.
A match's segment assignment is determined by its first served request with a segment assignment.

For each submitted request, we first determine if it can be assigned to a campus-bound match based on the agent's arrival time window. To do this, we filter the list of existing campus-bound matches to find those that are in the same or adjacent circle segment. We then sort the remaining matches by their similarity to the request (lexicographically with respect to the criteria \emph{same segment}, \emph{campus location already being serviced}, \emph{distance to closest home location of other agent}).
Afterward, we evaluate for the five most similar matches if the assignment of the request is feasible.

This is the case if

\begin{itemize}
    \item the resulting ride would fit between the vehicle’s rides before and after,
    \item the vehicle’s next ride still has to be feasible with this new ride’s resulting vehicle position and end time,
    \item no waiting times occur during the resulting ride,
    \item there exists a route in which said ride's agents arrive/depart during their time window, and
    \item all of said ride's agents spent no more minutes traveling than their accepted \emph{Ridepooling} travel time. 
\end{itemize}

Similar to our \emph{Ridesharing} mode, we perform the feasibility evaluations of our TSP-like problem using the respective helper functions.
From the feasible matches among the five evaluated, we select the one with the shortest total travel time. If these five matches do not yield at least one feasible solution, we continue to examine subsequent matches from our sorted list until we find a possible assignment.
If no such assignment is found, we investigate whether a new feasible match can be created with a free vehicle, i.e.~a vehicle which is not occupied during the request's time window.
For this, we sort all free vehicles based on their beeline distance to the agent's pickup location.
In the case of a found assignment, we nevertheless also investigate if a better match with a free vehicle is possible.

We choose the free vehicle over the assignment if the vehicle start position is closer (beeline distance) to the agent pickup location than all other pickup locations of the found assignment.
The chosen match is updated by adding the new agent to the passenger list and recalculating the necessary ride start time of this match.
Until this ride start further agents can join the match as long as the vehicle is not fully occupied.

The process is then repeated for the request's return trip demand. If no suitable match is found for this trip, the agent declines to partake in \emph{Ridepooling} and drives in his private vehicle instead.
Two new rides are created: one to campus and one back home, with the agent as the sole passenger. 
Whenever a ride start event occurs, the mode utilizes a helper function to create a finished ride object out of the match object and its precalculated optimal route.
These finished rides are used to evaluate the output metrics.
Finally, as an additional output our mode generates the output metrics for each minibus, summarized for the entire day.

\subsection{Demand Scenario Generation}\label{sec-demand}
The framework includes a default demand scenario for testing and comparing various mobility modes. This scenario represents students wishing to travel from home to the University of Würzburg campus and back. Although not directly integrated into the framework, we outline our process for generating this demand which is general enough to be applied to other regions.

The scenario combines two data sources: a list of students' home postcodes and faculty affiliations provided by the University of Würzburg
and data from the \emph{Mobilität in Deutschland (MiD) 2017} survey ('mobility in Germany 2017')~\cite{midSummary}. The latter was conducted by the German Federal Ministry for Digital and Transport, which questioned 316,361 individuals about their daily mobility patterns, including their modes of transport and trip timestamps. Notably, the MiD survey also includes students' responses regarding their arrival and staying times on campus.

Our approach to creating the demand scenario involves the following steps: We choose a set of campus locations, which represent the four distinct (sub)campuses in Würzburg, and assign faculties to these locations based on each faculty's main building address.
We define a demand region as postcode areas that are located (at least partly) within a radius of 55km around the faculty postcode regions since we deem it unlikely for students to live farther away.
This distance threshold aligns with the maximum daily commute of over 90\% of German commuters~\cite{pendler}.
Note that our dataset included home postcodes with significantly longer distances from campus. We assume that this is because some students provided their former home addresses at the time of answering the survey instead of their current Würzburg home addresses.
Therefore, we exclude these postcodes.
To generate a base population for our demand region with the expected amount of students per postcode,  
we utilize the tool OMOD~\cite{strobel2023omod} which creates realistic home locations for a given area based on MiD data.
We assign to each student a OMOD-generated home address within their postcode area as the origin for their journey to campus, with the campus location of their corresponding faculty as the destination.

We further develop a probability distribution function for arrival and departure times using MiD data and the programming language R. 
R's \texttt{density} function for estimating kernel densities and \texttt{approxfun} function for data point interpolation were crucial to this process.
With this probability distribution we are able to generate samples for preferred arrival times in 5-minute intervals (e.g., 08:00, 08:05, 08:10), ensuring that students only submit wishes for arrival for these specific times.
Afterwards, we randomly assign these arrival times to the created students.

Since the arrival time influences the time students spend at campus, we group the questioned students into quartiles based on their preferred arrival times and their corresponding staying times at campus.
For each quartile we separately determine a probability distribution for the staying times.
For example, students arriving at 08:30 are assigned a distribution favoring stays of 7 hours or more while those arriving at 18:00 are assigned a distribution favoring stays of 2 hours or less.
Based on an agent's arrival and staying time at campus we calculate their preferred departure time (arrival + staying = departure).
For the submission time of each request itself we draw from a uniform distribution between 30 and 360 minutes before each arrival time.

\section{Case Study} \label{sec:case_study}

In this chapter we present our case study design and results as well as a sensitivity analysis in which we investigate the effect of adapting our case study's configuration.

\subsection{Description of Evaluated Scenario} \label{sec-scenario}

In order to evaluate our implemented mobility modes, we utilize our demand scenario for the University of Würzburg, which we generated as described in Section~\ref{sec-demand}.
To ensure comparability, we input the same circumstances (demand and parameter configurations) for all evaluated modes.
We discuss deviations from the base case in form of a sensitivity analysis in Section~\ref{subsubsec:sensitivity}.

\begin{table}
    \centering
    \large
    \resizebox{0.65\textwidth}{!}{
        \begin{tabular}{ | m{7cm} || m{5cm} | }
            \hline
            & \\
            Parameter & Value \\
            & \\
            \hline\hline
            inner radius (km) & 2.0 \\
            \hline
            outer radius (km) & 25.0 \\
            \hline
            number of \emph{Ridepooling} vehicles & 250 \\
            \hline
            \emph{Ridepooling} vehicle seating capacity & 6 \\
            \hline
            student vehicle seating capacity & 5 \\
            \hline
            flexible time (min) & 15 \\
            \hline
            accepted walking distance (m) & 1200 \\
            \hline
            accepted \emph{Ridesharing} ride time & x + log1.4(x) \\
            &with direct travel time as x\\
            \hline
            accepted \emph{Ridepooling} ride time & x + log1.2(x) \\
            &with direct travel time as x\\
            \hline
            student vehicle fuel consumption (l/km) & 0.048 \\
            \hline
            student vehicle $CO_2$ emissions (g/km) & 126 \\
            \hline
            student vehicle fuel price (\euro/l) & 1.719 \\
            \hline
            \emph{Ridepooling} vehicle fuel consumption (l/km) & 0.038 \\
            \hline
            \emph{Ridepooling} vehicle $CO_2$ emissions (g/km) & 2578.95 \\
            \hline
            \emph{Ridepooling} vehicle fuel price (\euro/l) & 1.719 \\
            \hline
            stop time (min) & 1 \\
            \hline
            number of circle segments & 18 \\
            \hline
            segmentation exclusion radius (m) & 3000 \\
            \hline
        \end{tabular}}
    \caption{\centering{Parameter set for our evaluated scenario.}}
    \label{tab:baseCaseParameters}
\end{table}

\begin{figure}
    \centering
    \includegraphics[page=1,trim={0 1cm 0 0},clip,width=0.6\linewidth]{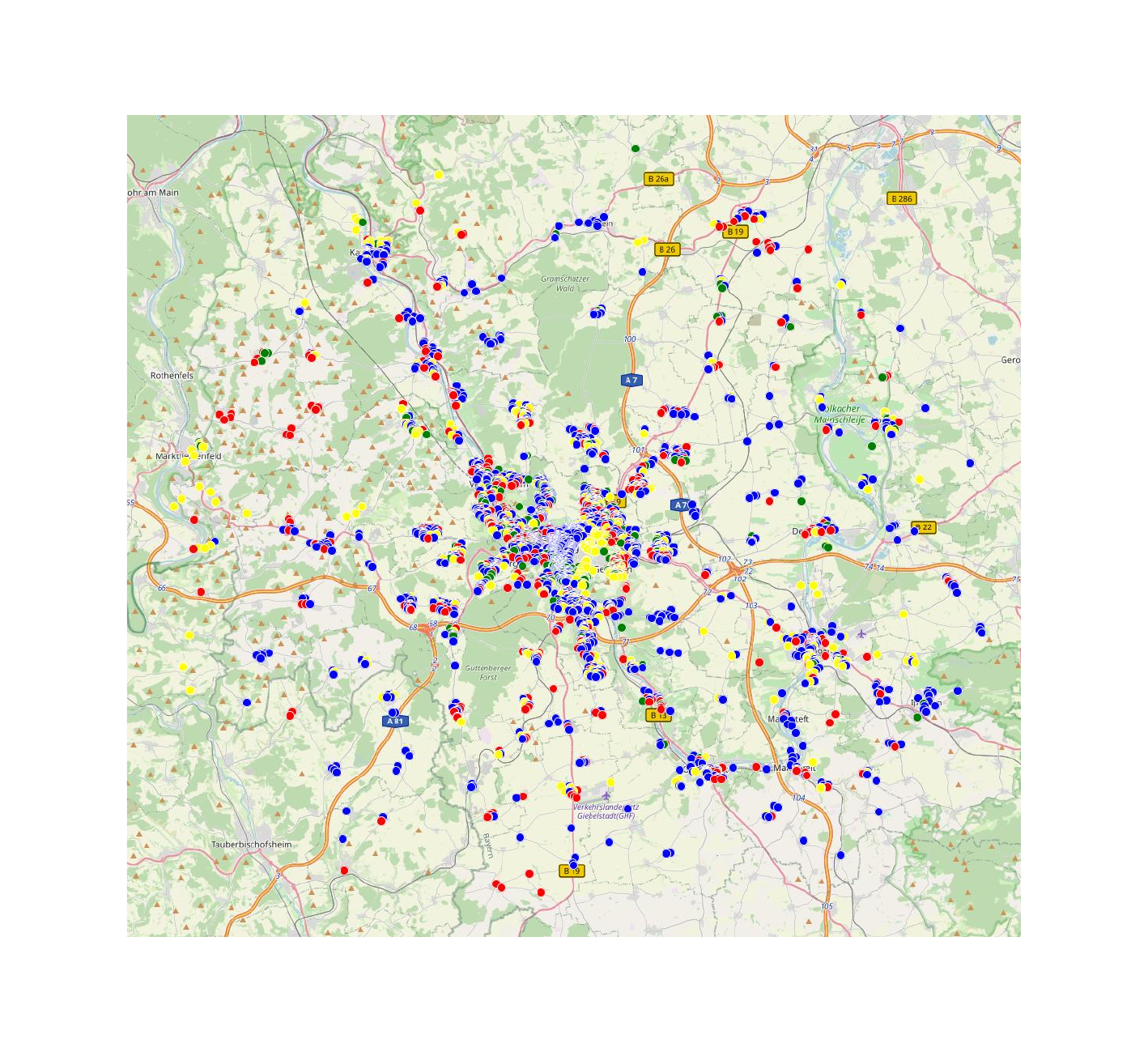}
    \caption{Agents in and around Würzburg (distance to campus in [2.0,25.0] meters). The colors reflect the different destination campus locations.}
    \label{fig:markers}
\end{figure}

To ensure a realistic scenario, we choose to serve only a subset of the generated demand. Specifically, we select students whose home locations are within a reasonable distance from their campus location, i.e. a distance sensible for utilizing mobility alternatives. For example, expecting minibuses to cover large distances could result in inefficient use of minibuses and excessive empty travel times.
Thus, we create a more feasible scenario for our proposed mobility options by setting our \texttt{outer radius} configuration parameter to 25.0~km.
To exclude students who live particularly close to campus, we set the \texttt{inner radius} to 2.0~km. This threshold reflects that, according to the MiD~\cite{mid},
people aged 20 to 30 prefer micromobility choices such as walking or cycling for short distances, with about 80\% choosing these forms of transportation.
Figure~\ref{fig:markers} depicts the home locations of our generated students who live in this radius which is a total of 6524 agents. The markers are colored based on each student's campus destination location.

The full set of input parameters used in our base case is shown in Table~\ref{tab:baseCaseParameters}.
We set a maximum \texttt{accepted walking distance} of 1200 meters for each agent which aligns with the assertion of German federal ministries that distances up to this are walkable~\cite{haltestellen}. 

To allow for flexibility in student arrival and departure times, we set the \texttt{flexible time} to 15 minutes, creating a time window of 30 minutes. This approach considers that students are willing to arrive or depart slightly earlier or later than their preferred times.
We calculate each agent's \texttt{accepted ride time} for both alternative mobility modes using a logarithmic relationship between their direct car travel time (i.e.~the travel time that the student would have if traveling in their own car) and the acceptable detour time for alternative modes. As the direct ride time increases, the total allowed detour time also increases, but at a decreasing rate. For example, a student with a five-minute direct travel time might accept a ten-minute ride using an alternative mode, while a student with a thirty-minute direct time would not be willing to spend sixty minutes traveling.
For both \emph{Ridesharing} and \emph{Ridepooling}, we set the duration of a stop~(\texttt{stop time}) during a ride to one minute which accounts for passengers stowing luggage and choosing a place to sit.

For modeling the student vehicles we choose a slightly older model and thus use the VW Golf 2.0 TDI (year of
manufacture 2010)~\cite{studentauto}.
Since our \emph{Ridepooling} approach should strive for energy efficiency, we pick the Citroen Grand C4 Picasso (year of manufacture 2018) for modeling the \emph{Ridepooling} vehicles. This car model is a minibus that uses less diesel fuel than the student vehicle for the same distance.\cite{ridepoolingauto}

\begin{table}
    \large
    \centering
    \resizebox{\textwidth}{!}{
        \begin{tabular}{ | m{6.5cm} || m{3cm} | m{3cm} | m{3cm} | }
            \hline
            &&&\\
             & \emph{EverybodyDrives} & \emph{Ridesharing} & \emph{Ridepooling} \\
            &&&\\
            \hline\hline
            Total driven distance (km) & 120103.4 (100\%) & 62.65\% *\newline 57.80\% ** & 102.51\% \\
            \hline
            Total driven time (min) & 130833 (100\%) & 60.12\% *\newline 55.23\% ** & 114.09\% \\
            \hline
            Total CO$_2$ emissions (kg) & 15133 (100\%) & 48.94\% *\newline 44.08\% ** & 81.44\% \\
            \hline
            Total fuel costs (\euro) & 9909.98 (100\%) & 61.73\% *\newline 44.08\% ** & 82.76\% \\
            \hline
            Total empty covered distances (km) & 0 & 0 & 45792.57 \\
            \hline
            Number of lost students & 0 & 628 & 0 \\
            \hline
            Number of driving students & 6524 & 2841 & 449 \\
            \hline
            Number of rides & 13048 & 6310 *\newline5682 **  & 4620 \\
            \hline
            Number of rides with $>1$ students & 0 & 2878 & 3176 \\
            \hline
            Avg. CO$_2$ emissions per student (kg) & 2.32 & 1.14 *\newline1.02 ** & 1.89 \\
            \hline
            Avg. CO$_2$ emissions per ride (kg) & 1.16& 1.17 *\newline1.17 ** & 2.53 \\
            \hline
            Avg. fuel cost per student (\euro) & 1.52 & 0.94 *\newline0.67 **& 1.26 \\
            \hline
            Avg. fuel cost per ride (\euro) & 0.76 & 0.97 *\newline0.77 ** & 1.68 \\
            \hline
            Avg. seat occupancy & 1 & 2.07 *\newline2.30 **& 2.82 \\
            \hline
        \end{tabular}}
    \caption{\centering{The output metrics for the mobility modes accumulated for all rides of one day. * including multi-station carsharing rides home. ** excluding multi-station carsharing rides home.}}
    \label{tab:outputmetrics}
\end{table}

\subsection{Case Study Results}
In this section we present and compare the results for the three modes, \emph{EverybodyDrives}, \emph{Ridesharing} and \emph{Ridepooling}, which we each evaluate for one demand scenario (see Section \ref{sec-scenario}).

The aggregated results are shown in Table~\ref{tab:outputmetrics}.
The total driven distance accumulates all routes traveled by each vehicle in the mobility mode scenario.
This metric, along with total time, CO$_2$ emissions and fuel costs, allows for a conclusive performance assessment. To better demonstrate the impact of our alternative mobility modes, we present the main metrics' results as percentages relative to the \emph{EverybodyDrives} baseline.

\emph{Empty covered distance} only pertains to \emph{Ridepooling} and denotes the routes covered by \emph{Ridepooling} vehicles without any students occupying seats.
Similarly, \emph{Number of lost students} exclusively applies to \emph{Ridesharing} and refers to students without a \emph{Ridesharing} opportunity for riding home.
Moreover, the \emph{Ridesharing} column in Table~\ref{tab:outputmetrics} frequently displays two values, based on two different assumptions regarding lost students. For the first numbers that are shown, the numbers incorporate a car drive back for these students, which models, e.g., that the students make use of multi-station carsharing for their journeys home, while the second excludes the trips home them under the assumption that they utilize public transportation to travel home.
For CO$_2$ emissions and fuel costs we report averages per student and per ride (where a ride constitutes a vehicle’s complete route for transporting a subset of students from starting point to the last drop-off destination).
Moreover, the average seat occupancy denotes the average number of students per ride across all trips.
This includes single-driver rides where alternative mobility modes are not utilized.

For the chosen parameters both \emph{Ridesharing} and \emph{Ridepooling} lead to reductions of the total daily fuel costs and CO$_2$ emissions under our modeling assumptions.
As to \emph{Ridesharing}, this is due to less total vehicle kilometers being driven due to a higher seat occupancy and therefore less rides in total compared to \emph{EverybodyDrives}.
\emph{Ridepooling}, however, leads to an increase of total traveled time and distances since the autonomous vehicles need to cover additional distance without any passengers in order to reach pickup destinations.
Nevertheless, under the assumption that the \emph{Ridepooling} vehicles are more energy efficient than the average student vehicle, fewer liters of diesel fuel are being used for the same distances, resulting in less emissions and fuel costs overall.

\begin{figure}
        \begin{subfigure}[c]{0.47\linewidth}
        \includegraphics[page=1,width=\linewidth]{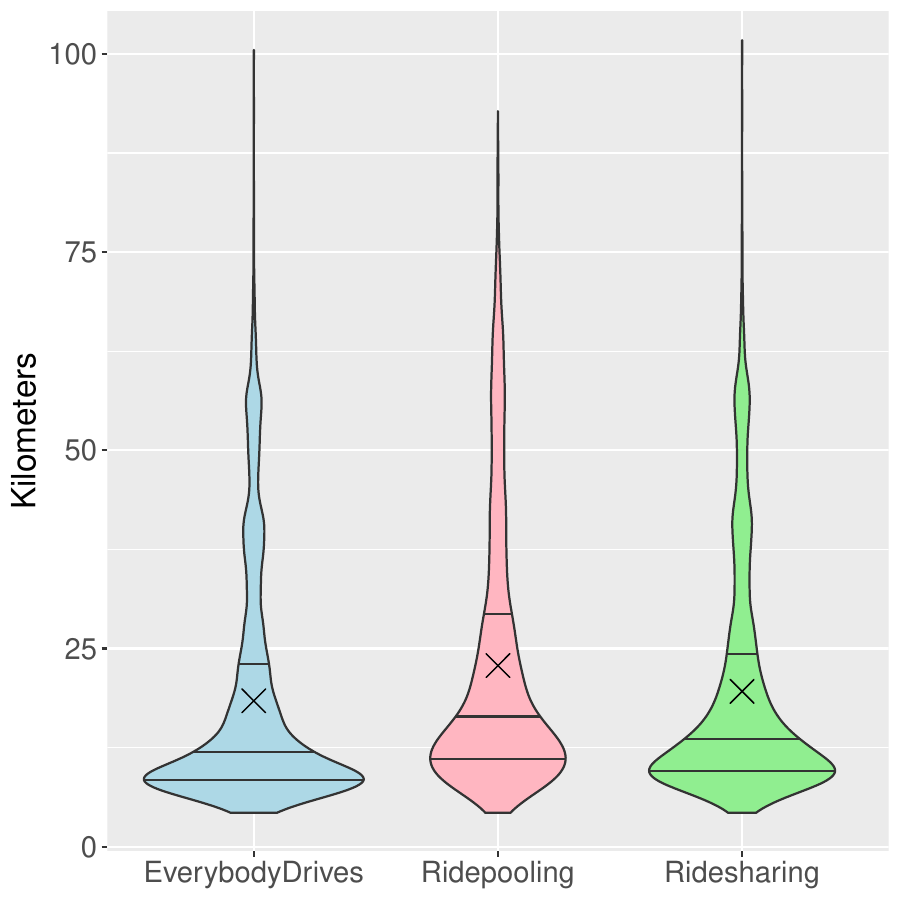}
        \caption{Daily distance per agent.}
    \end{subfigure}
    \hfill
    \begin{subfigure}[c]{0.47\linewidth}
        \includegraphics[page=1,width=\linewidth]{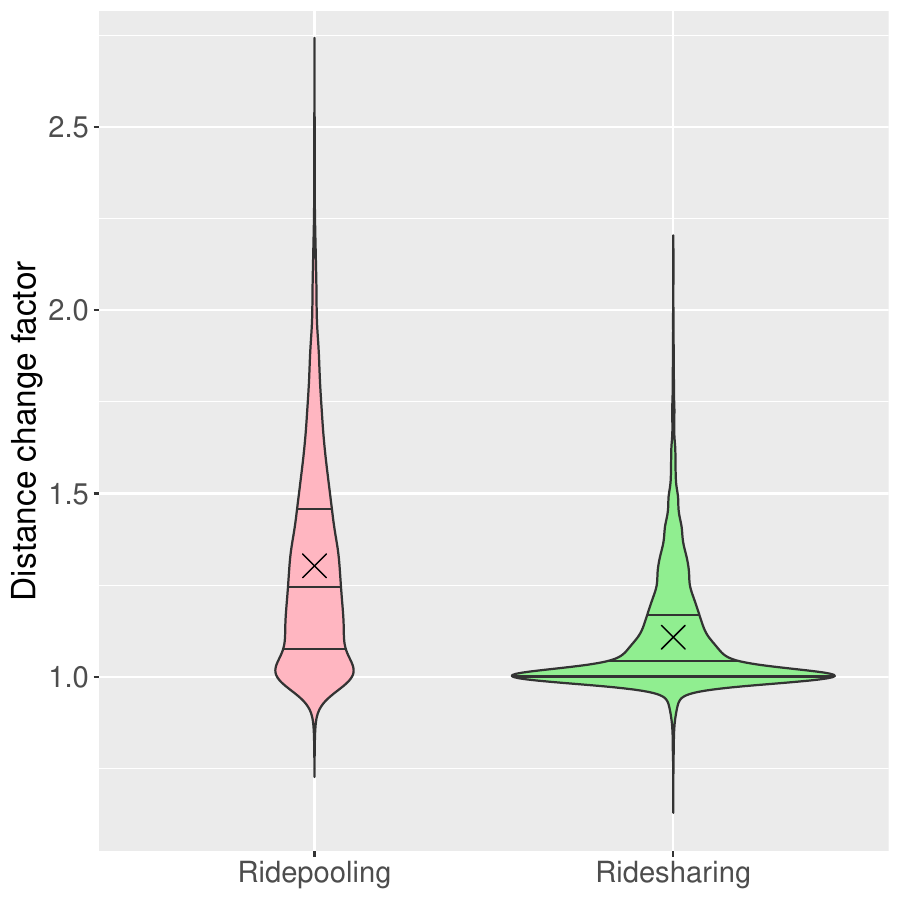}
        \caption{Distance change factor.}
    \end{subfigure}
    \caption{Effect of modes on distances traveled by agents (including walking distance in \emph{Ridesharing}).}
    \label{fig:gesamtkm}
\end{figure}

Furthermore, we examine the effect of the alternative modes on the individual agents and display these results in the form of violin plots.
These plots visualize the distribution of the examined  quantities. The lines inside the plots indicate median and quartiles, the cross indicates the mean.

As we can see in Figure~\ref{fig:gesamtkm}(a), the average travel time does not change much for both \emph{Ridesharing} and \emph{Ridepooling} compared to the baseline mode~\emph{EverybodyDrives}.
However, because of picking up or dropping off other passengers, \emph{Ridesharing} and \emph{Ridepooling} lead to longer distances on average.
\emph{Ridepooling} more so, since \emph{Ridepooling} passengers accept slightly longer detours in our scenario (due to being picked up directly at their homes).
Figure~\ref{fig:gesamtkm}(b) shows the relative change in travel distance which we demonstrate by dividing each agent's daily traveled distance for an alternative mode by their direct daily distance traveled (i.e.~the agent's daily~\emph{EverybodyDrives} travel distance). We can see that, on average, agents travel slightly longer when using either alternative mode.
Furthermore, for both \emph{Ridepooling} and \emph{Ridesharing} we see a few cases where agents travel shorter distances compared to the \emph{EverybodyDrives} mode. For the \emph{Ridepooling} mode, this can happen when there is a less direct but faster route available, e.g., a highway, which is used when a student drives by himself but is not used by the \emph{Ridepooling} vehicles who pick up more students. For \emph{Ridesharing}, this can happen when walking to the driver presents a shortcut for passengers.

\begin{figure}
    \begin{subfigure}[c]{0.47\linewidth}
        \includegraphics[page=2,width=\linewidth]{pictures/withWalkingFigures.pdf}
        \caption{Daily minutes per agent.}
    \end{subfigure}%
    \hfill
    \begin{subfigure}[c]{0.47\linewidth}
        \includegraphics[page=2,width=\linewidth]{pictures/factorFiguresWithWalking.pdf}
        \caption{Time change factor.}
    \end{subfigure}%
    \newline
    \begin{subfigure}[c]{0.47\linewidth}
        \includegraphics[page=1,width=\linewidth]{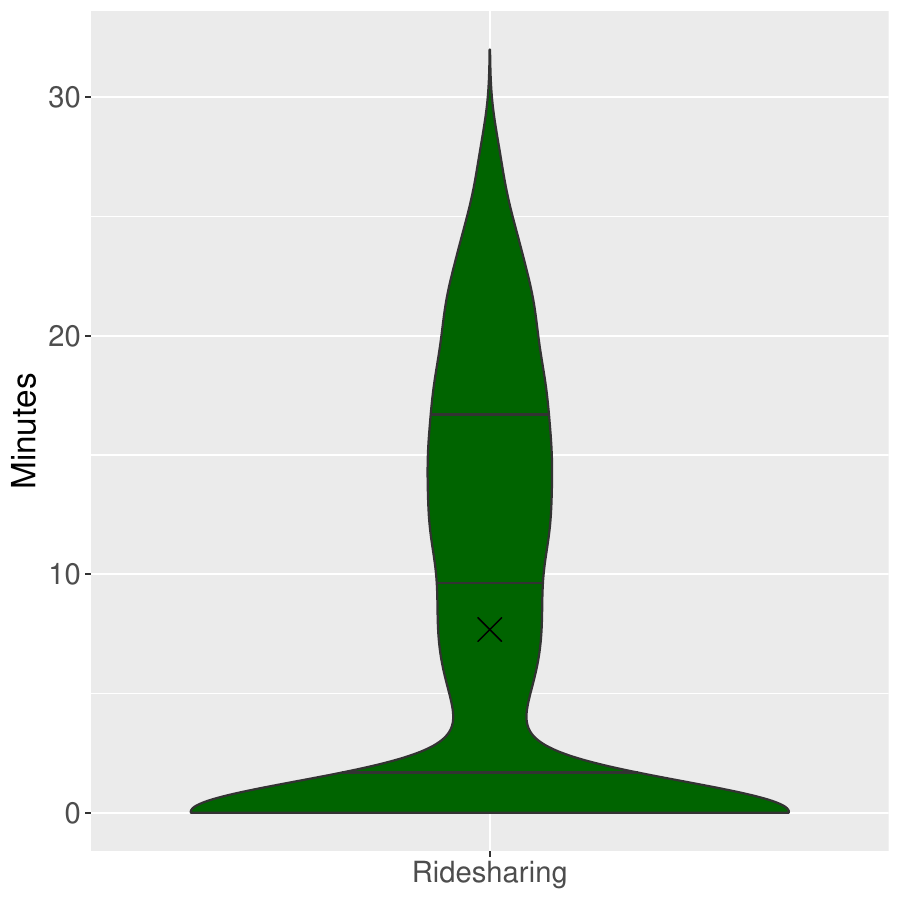}
        \caption{Daily walking time per agent.}
    \end{subfigure}%
    \hfill
    \begin{subfigure}[c]{0.47\linewidth}
        \includegraphics[page=3,width=\linewidth]{pictures/walkingFigures}
        \caption{Percentage of walking time of total travel time.}
    \end{subfigure}%
    \caption{Effect of modes on agents' travel time (including walking time).}
    \label{fig:gesamtminuten}
\end{figure}

Similar to the increase in traveled distances, agents who use \emph{Ridesharing} and \emph{Ridepooling} also spend more time traveling on average because of detours. The detours have a bigger impact on travel time than on distance because picking up and dropping off other passengers takes additional time (one minute in our model).

We further analyze how much time agents spend walking in \emph{Ridesharing} (see Figure~\ref{fig:gesamtminuten}(c)).
Most agents don't walk at all, either because they drive themselves or live in the same building as the driver.
The majority of passengers walk less than ten minutes daily.
However, Figure~\ref{fig:gesamtminuten}(d) reveals that many agents spend a considerable portion of their total travel time walking.
To address this, future versions of the \emph{Ridesharing} model could include walking time for the calculation of the acceptable ride time.

\begin{figure}
    \begin{subfigure}[c]{0.47\linewidth}
        \includegraphics[page=3,width=\linewidth]{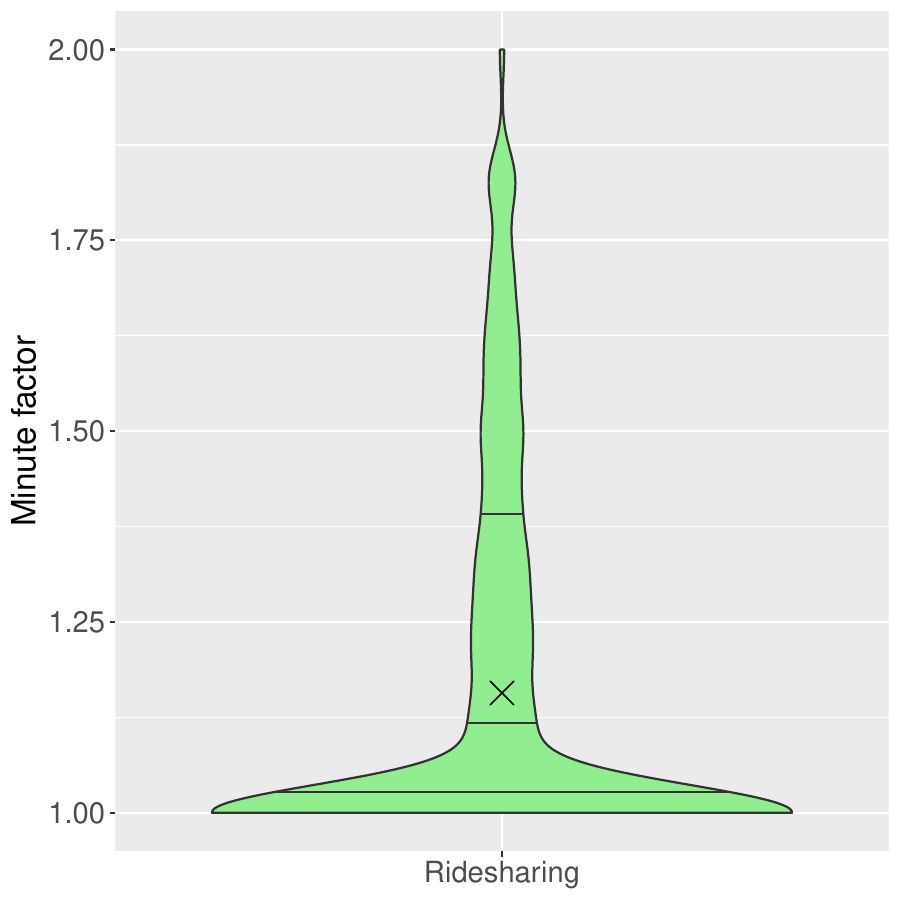}
        \caption{Time change factor of \emph{Ridesharing} drivers (not traveling alone).}
    \end{subfigure}
    \hfill
    \begin{subfigure}[c]{0.47\linewidth}
    \includegraphics[page=4,width=\linewidth]{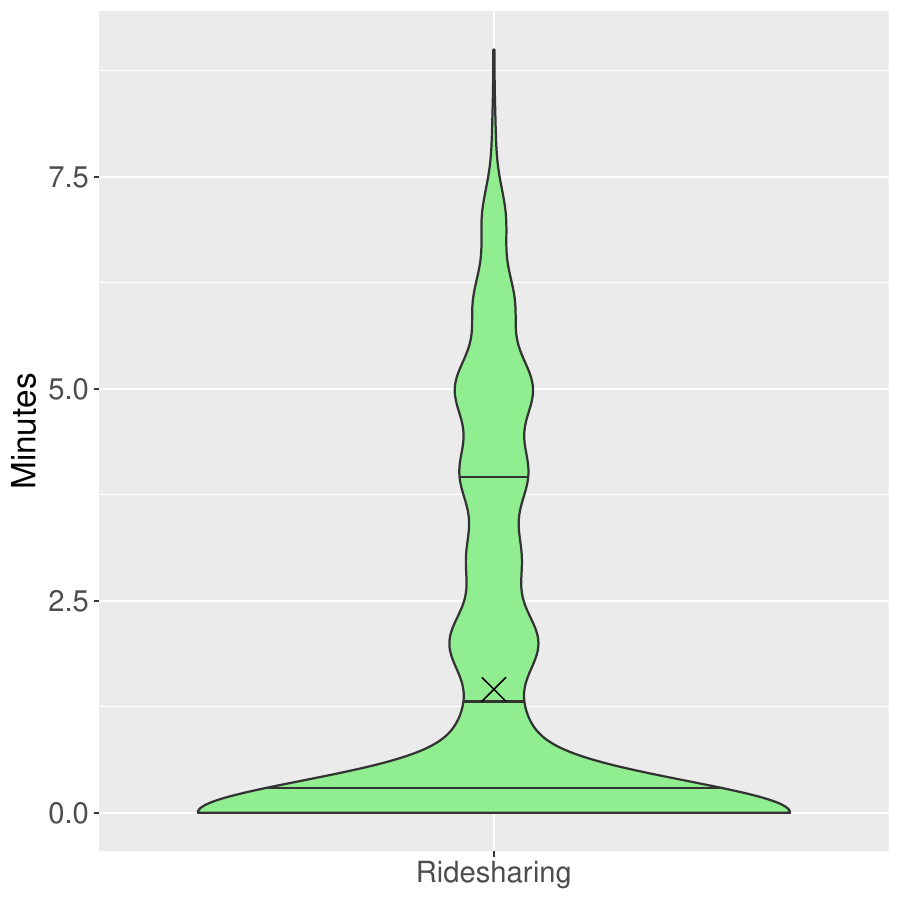}
        \caption{Additional travel time for \emph{Ridesharing} drivers (not traveling alone) per ride.}
    \end{subfigure}
    \caption{\emph{Ridesharing} effects on drivers (not traveling alone) regarding travel time.}
    \label{fig:extraminutes}
\end{figure}

Drivers experience less inconvenience from \emph{Ridesharing} than the average agent based on the time change factor (relative change in travel time between baseline and alternative modes), as evidenced by their smaller average time change factor (compare Figures~\ref{fig:gesamtminuten}(b) and~\ref{fig:extraminutes}(a)).
On average, a driver with passengers only experiences an additional travel time of 1.45 minutes per ride.

\subsection{Sensitivity Analysis} \label{subsubsec:sensitivity}
In this chapter we discuss our sensitivity analysis regarding the influence of a number of selected parameters on our output metrics. 
Our results indicate that, generally, the flexibility of the time windows for arrival and departure windows has a significant effect on the daily amount of rides since these time windows are crucial for successfully matching requests.

\begin{figure}
    \begin{subfigure}{0.48\linewidth}
        \includegraphics[page=89,width=\textwidth]{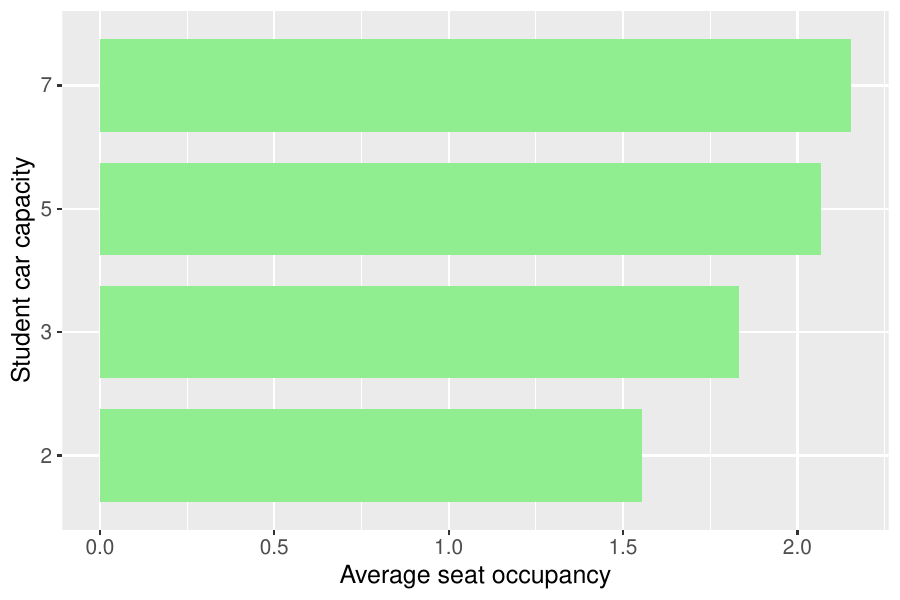}
    \end{subfigure}
    \hfill
    \begin{subfigure}{0.48\linewidth}
        \includegraphics[page=110,width=\textwidth]{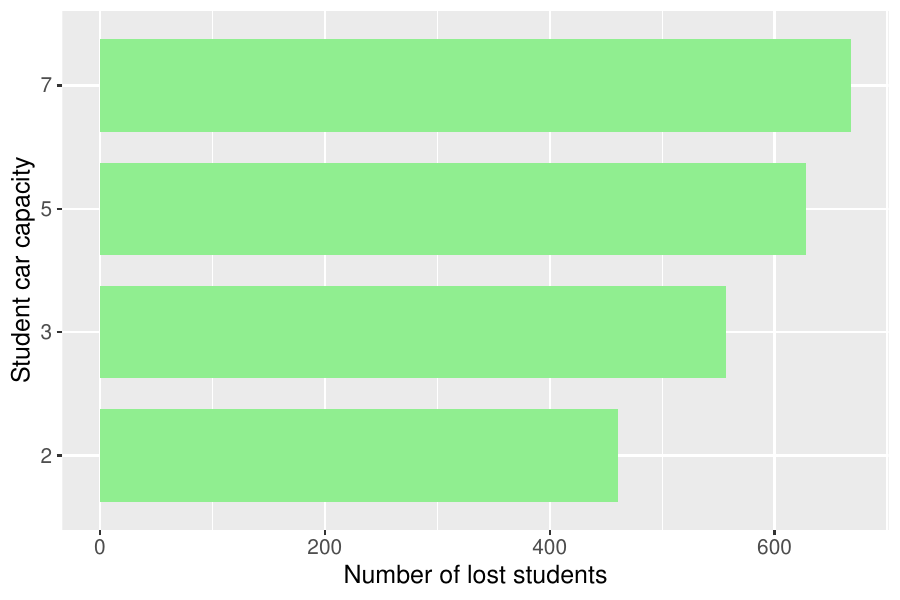}
    \end{subfigure}%

    \vspace{0.7cm}
    \begin{subfigure}{0.48\linewidth}
        \includegraphics[page=91,width=\textwidth]{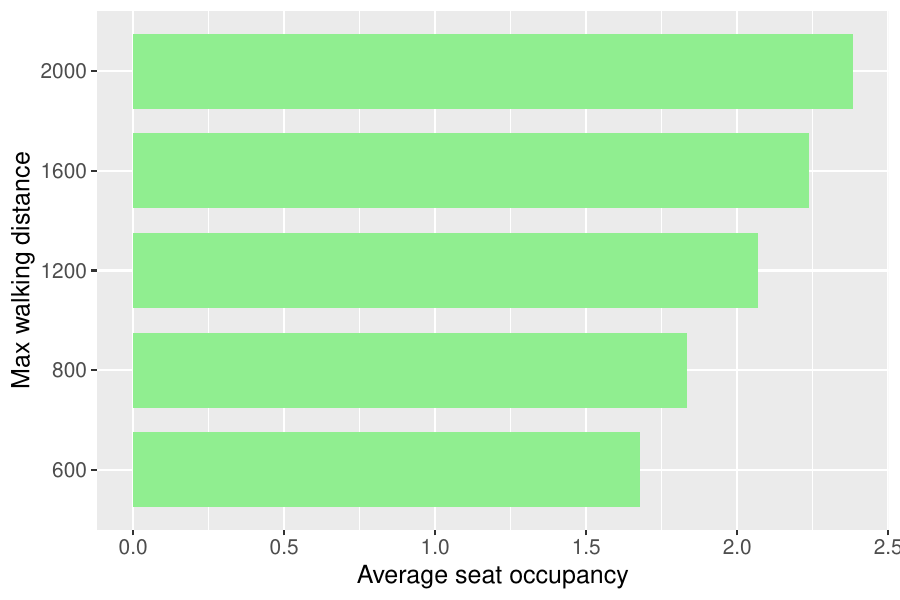}
    \end{subfigure}
    \hfill
    \begin{subfigure}{0.48\linewidth}
        \includegraphics[page=112,width=\textwidth]{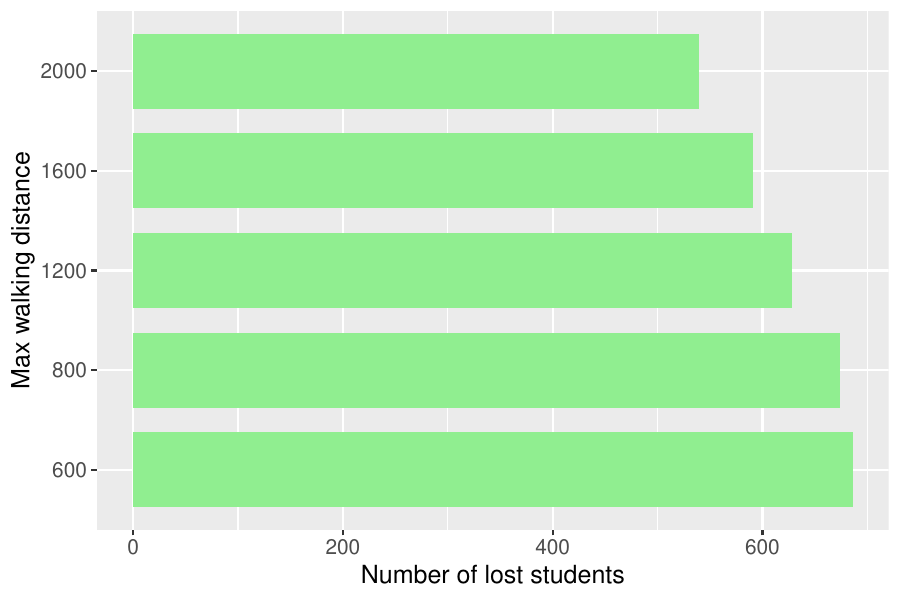}
    \end{subfigure}
    \caption{Effect of differing input parameter values on output metrics for \emph{Ridesharing}.}
\label{fig:walkingSensitivity}
\end{figure}

Some of our input parameters only influence \emph{Ridesharing}, namely the number of seats of the student vehicle and the distance an agent is willing to walk to or from the driver’s home.
In the top of Figure~\ref{fig:walkingSensitivity} we can see the effect of changing the number of available seats in the student car on the average seat occupancy and the number of lost students.
Interestingly, increasing the number of available seats in student cars raises the average ride occupancy but also leads to more lost students. This occurs because more agents traveling as passengers (resulting in a higher average seat occupancy) means less available cars for return trips which leads to more lost students because not everyone finds a ride back at the desired return time.
On the other hand, a larger accepted walking distance increases average seat occupancy, as well, while also reducing the number of lost students. This suggests that the accepted walking distance is crucial for the success of our \emph{Ridesharing} approach. However, expecting students to walk longer distances might significantly decrease potential passengers' acceptance of \emph{Ridesharing}.

\begin{figure}
    \begin{subfigure}{0.48\linewidth}
        \includegraphics[page=85,width=\textwidth]{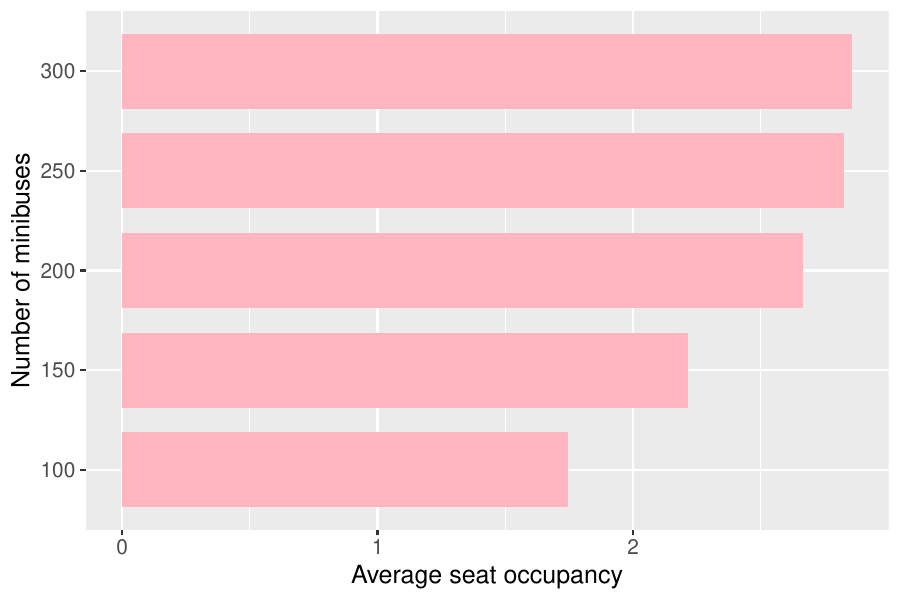}
    \end{subfigure}
    \hfill
    \begin{subfigure}{0.48\linewidth}
        \includegraphics[page=113,width=\textwidth]{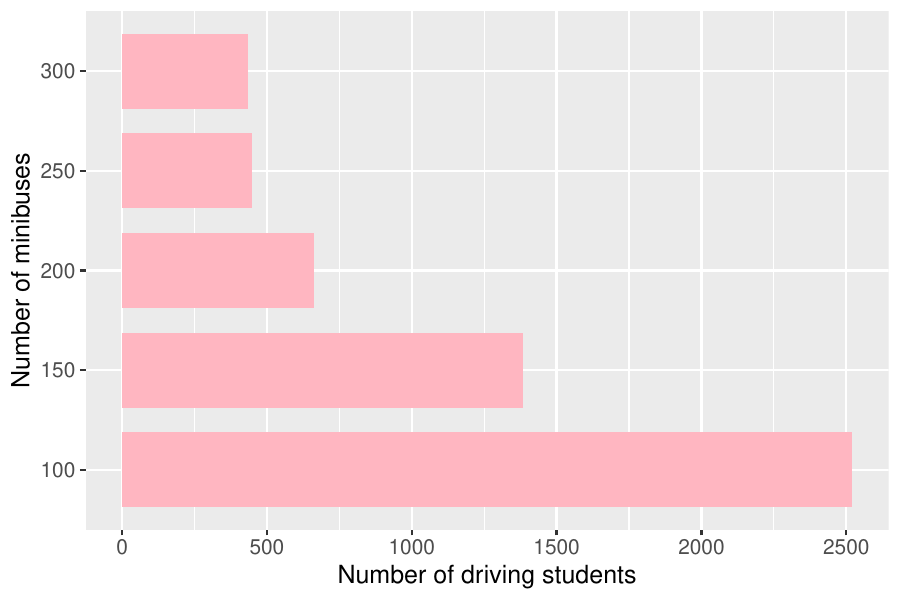}
    \end{subfigure}%

    \vspace{0.7cm}
    \centering
    \begin{subfigure}{0.48\linewidth}
    \includegraphics[page=86,width=\textwidth]{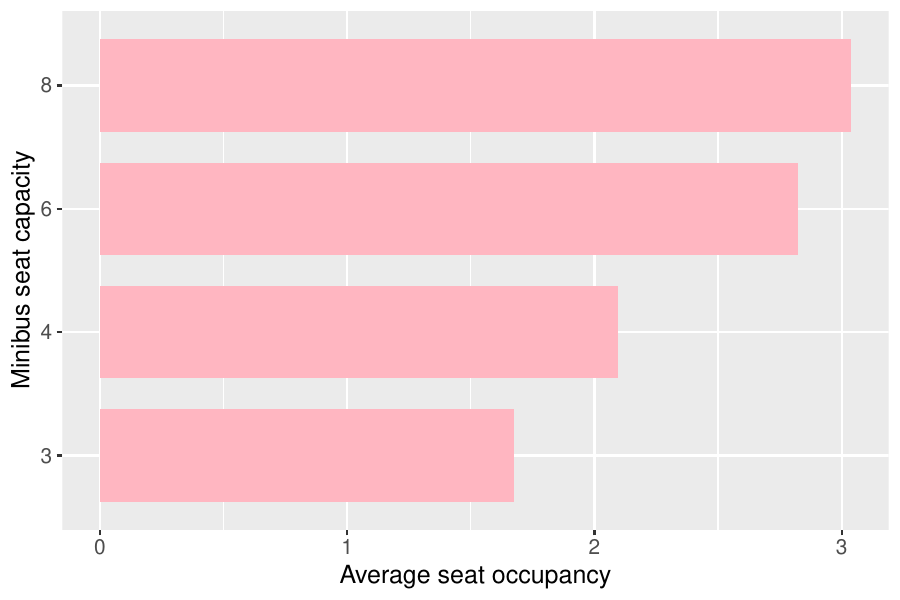}
    \end{subfigure}
    \vspace{0.5cm}
    \caption{Effect of differing input parameter values on output metrics for\emph{Ridepooling}.}
    \label{fig:buscountSensitivity}
\end{figure}

Furthermore, among the mode-specific parameters for \emph{Ridepooling}, we investigate the impact of the fleet size (number of available \emph{Ridepooling} vehicles) and the amount of available seats in the vehicle.
As shown at the top of Figure~\ref{fig:buscountSensitivity}, a smaller \emph{Ridepooling} fleet results in significantly more students driving their own cars. This is due to the limited vehicle fleet reaching capacity which prevents additional matches.
However, while a smaller fleet reduces average seat occupancy, the change is less dramatic than expected considering the increase in self-driving students.
We attribute this to many students who drive themselves when the fleet is small becoming sole passengers in \emph{Ridepooling} vehicles when the fleet is larger.
As a result, CO$_2$ emissions in our scenario are not significantly affected by fleet size changes.
We conclude that the fleet size can be reduced (e.g, for the goal of cost savings) without substantially increasing emissions.

The influence of the number of available seats in a \emph{Ridepooling} vehicle is demonstrated at the bottom of
Figure~\ref{fig:buscountSensitivity}. Changing the seating capacity significantly affects the average daily seat occupancy and, as a result, CO$_2$ emissions, with more seats resulting in lower emissions. This suggests that using vehicles with an appropriate number of seats could enhance the sustainability of \emph{Ridepooling}.
However, it is important to note that vehicles with more seats are typically larger and heavier which potentially leads to higher emissions~\cite{heavyvehicles}.
This trade-off should be considered for choosing \emph{Ridepooling} vehicles.

\section{Conclusions and Further Research}~\label{sec:conclusion}

In this paper we have developed an extendable framework that allows to evaluate and compare different mobility modes for campus mobility.
While the framework already includes three default mobility modes and a default demand scenario, these modules are interchangeable, allowing the user to test the implemented mobility modes on a demand scenario of their choice, or to include and compare own mobility modes.

We conducted a case study using the mobility modes \emph{EverybodyDrives}, \emph{Ridesharing}, and \emph{Ridepooling} on a scenario simulating students commuting to the Würzburg university campus. We assess the performance of these modes using sustainability and service quality metrics, including daily $CO_2$ emissions and travel time. We also carried out a sensitivity analysis to recognize which input parameters and constraints influence the results of each mode in our scenario the most.
Our results indicate that in the evaluated scenario, where all vehicles use diesel fuel, \emph{Ridesharing} produces the lowest $CO_2$ emissions due to a significant reduction in total distance traveled. 

\emph{Ridepooling}, however, did not achieve a reduction in distance traveled because vehicles have to make empty trips to pick up students and return to the depot. As a result, the sustainability of \emph{Ridepooling} depends entirely on the chosen vehicle type. In our scenario, we used a \emph{Ridepooling} vehicle with slightly lower emissions per kilometer than the student car model which led to small reductions for both $CO_2$ emissions and fuel costs.
\emph{Ridesharing} shows a higher rate of $CO_2$ reduction compared to cost reduction since approximately 10\% of participants cannot find suitable \emph{Ridesharing} matches for their return trips, forcing them to use the scenario's more expensive alternative. Despite this, average fuel costs still decrease with \emph{Ridesharing}.
With regard to travel time, \emph{Ridepooling} performs worse due to a considerable increase in average travel minutes caused by additional pick-up stops.

In conclusion, our approaches for both \emph{Ridesharing} and \emph{Ridepooling} successfully reduce $CO_2$ emissions and costs while observing students' quality of service constraints. For our specific scenario, \emph{Ridesharing} outperforms \emph{Ridepooling} in regards to average emissions, costs, traveled distance, and time. However, these results highly depend on several factors, such as students' willingness to walk, the vehicle seating capacity or the emissions per kilometer of the employed vehicles.
Furthermore, our \emph{Ridesharing} approach leads to about 10\% of students lacking a return trip. 
This shows the need to think about alternative means of transport for this students in order to make this an acceptable alternative.

There is no definitive answer as to which mobility mode is best for campus mobility. We recommend that universities or institutions that aim to improve the sustainability of their campus mobility conduct a comprehensive survey of their students' travel demands and requirements.
This will help with identifying critical quality of service metrics. For instance, while our \emph{Ridepooling} approach results in longer travel times, it offers a home-pickup service. On the other hand, our \emph{Ridesharing} approach requires students to walk to meet their ride, up to a certain distance. Student preferences might vary, e.g.,~some students would rather avoid any walking, even if it results in longer vehicle travel times. These preferences could also differ between universities. Thus, a tailored approach regarding institutional needs and student preferences is essential.
We additionally regard it as crucial for universities to ensure that all students have alternative options in case of not finding matches for their home ride.
Regarding our evaluated scenario, based on our results we would advise policy makers to
consider employing \emph{ridepooling} with electric vehicles that, preferably, use ecologically
sustainable energy sources since this leads to the least amount of emissions and ensures
that students can get a ride home.

Our evaluation is based on a simplified scenario that excludes traffic influences and unexpected incidents such as students being late for pickup. 
It would be interesting to investigate to what extent this uncertainty has an impact on the relative performance of the considered mobility modes.
For instance, both the impact of traffic on the feasibility and outcomes of these modes as well as, in turn, the effect of each mode on traffic and congestion are critical for decision-making.
Moreover, in our current model we assume the same quality of service constraints for all agents as well as identical attributes for all vehicles (of the same type).
Future research could investigate the effects of differing agent demands and vehicle attributes, potentially incorporating probabilistic distributions.
Additionally, cost factors like acquisition costs, service fees, and vehicle wear could be considered in future work.
Finally, examining the impact of necessary refueling stops and the logistics of employing drivers for \emph{ridepooling}, including shift management, could aid in receiving more realistic results and promoting alternative campus mobility modes.

\section{Statements and Declarations}
The authors have no competing interests to declare that are relevant to the content of this article.

\bibliography{sn-bibliography}

\end{document}